\documentclass[aps,prd,letterpaper,superscriptaddress,groupedaddress,preprintnumbers,floatfix,twocolumn,nofootinbib]{revtex4}  
\usepackage{graphicx,comment}  % needed for figures
\usepackage{dcolumn}   % needed for some tables
\usepackage{bm}        % for math
\usepackage{amssymb}   % for math

\usepackage{subfigure}
\usepackage{multirow}
\usepackage{slashed}
\usepackage{placeins}
\usepackage{epstopdf}
\usepackage{mathtools}
\usepackage{color}
\usepackage{tikz,pgfplots,filecontents}
\pgfplotsset{compat=newest}

\usepackage{enumerate}
\usepackage{setspace}
\usepackage{amsmath}
\usepackage{bm}               
\usepackage{amssymb}
\usepackage{amsfonts}
\usepackage{mathtools}
\usepackage{dsfont}
\usepackage{commath}
\usepackage{physics}
\usepackage{tensor}
\usepackage{amsfonts}    
\usepackage{amsmath}
\usepackage{dcolumn}\usepackage{tabularx}
\usepackage{listings}

\usepackage{algorithm}
\usepackage[compatible]{algpseudocode}
\renewcommand{\algorithmiccomment}[1]{\bgroup\hfill//~#1\egroup}

\definecolor{c1}{RGB}{206,0,0}
\definecolor{c2}{RGB}{249,149,0}
\definecolor{c3}{RGB}{153,0,210}
\definecolor{c4}{RGB}{0,109,219}
\definecolor{c5}{RGB}{0,146,146}
\definecolor{c6}{RGB}{255,109,182}

\pgfplotscreateplotcyclelist{cycle1}{
{c1}, {c2}, {c3}, {c4}, {c5}, {c6}
}

\begin{document}

\preprint{
\vbox{
\hbox{MIT-CTP/5138}
\hbox{ADP-19-17/T1097}
}}

\title{Accelerating lattice quantum field theory calculations via interpolator optimization using NISQ-era quantum computing}
\author{A.~Avkhadiev}\affiliation{Center for Theoretical Physics, Massachusetts Institute of Technology, Cambridge, MA 02139, U.S.A.}\affiliation{Perimeter Institute for Theoretical Physics, Waterloo, Ontario N2L 2Y5, Canada}
\author{P.~E.~Shanahan}\affiliation{Center for Theoretical Physics, Massachusetts Institute of Technology, Cambridge, MA 02139, U.S.A.}
\affiliation{Perimeter Institute for Theoretical Physics, Waterloo, Ontario N2L 2Y5, Canada}
\author{R.~D.~Young}\affiliation{CSSM, Department of Physics,
  University of Adelaide, Adelaide SA 5005, Australia}

\begin{abstract}
The only known way to study quantum field theories in non-perturbative regimes is using numerical calculations regulated on discrete space-time lattices.
Such computations, however, are often faced with exponential signal-to-noise challenges that render key physics studies untenable even with next generation classical computing.
Here, a method is presented by which the output of small-scale quantum computations on Noisy Intermediate-Scale Quantum era hardware can be used to accelerate larger-scale classical field theory calculations through the construction of {\it optimized interpolating operators}. The method is implemented and studied in the context of the 1+1-dimensional Schwinger model, a simple field theory which shares key features with the standard model of nuclear and particle physics.
\end{abstract}

\maketitle

Numerical approaches to quantum field theory are the only known way to make predictions for a wide range of physical quantities from the standard model of particle physics, our best current theory of nature at the smallest scales. Standard model calculations of nuclear physics processes---such as those needed to interpret experiments using nuclei as targets---are particularly challenging. In particular, the strong-interaction component of the standard model, which is encoded in the theory of quantum chromodynamics (QCD), can not be approached analytically at the relevant energy scales. The only first-principles approach to QCD at these scales is numerical: a discretized form of the QCD equations can be solved using supercomputers through Monte Carlo integration on a finite four-dimensional grid representing space-time \cite{Wilson:1974sk,Gattringer:2010zz}. 
This technique, named lattice quantum field theory (LQFT), plays an important role in modern particle and nuclear physics and has been essential in testing the standard model against precise measurements of the decays and interactions of particles at frontier machines such as the Large Hadron Collider~\cite{Aoki:2019cca,Lehner:2019wvv}. 
Calculations of nuclei, however, are limited by exponentially bad scaling of computational cost with the atomic number of the system being studied. Using current methods, direct studies of nuclei with tens of nucleons, as relevant to diverse physics programs from direct searches for dark matter to neutrino physics, will remain intractable, even with the advent of exascale classical computing in the next years; progress on this front will require a revolutionary approach, and there is great interest in the potential applications of quantum computing to overcome this challenge~\cite{Joo2019StatusBeyond,CarlsonQuantumPhysics}. Hybrid methods coupling classical and quantum computing offer a natural pathway to exploit quantum computation despite the small number of qubits, sparse qubit connectivity, lack of error-correction, and noisy quantum gates that are hallmarks of current and near-term quantum computing in the Noisy Intermediate-Scale Quantum (NISQ) era \cite{Preskill2018}. 

A significant contribution to the computational cost of LQFT studies could be eliminated by the construction of {\it optimized interpolating operators}, corresponding in broad terms to approximations to the quantum wavefunction of the desired state.
Precisely, to determine matrix elements of interest in some state in a LQFT computation, such as those describing an interaction or decay process, correlation functions are calculated which encode the creation, interaction, and annihilation, of the state in question. These correlation functions, however, receive contaminating contributions from the many other states with the quantum numbers of the state of interest. In order to reliably extract the desired piece, the contributions from all of these unwanted higher-energy states must be suppressed. 
Typically, this is achieved via an evolution in the Euclidean time of the calculation; the unwanted states are exponentially suppressed by the energy gap to the ground state at large times, but at the cost of an exponential growth in the statistical noise of the Monte Carlo sampling used in the computation (and thus computational cost). 
By using optimized interpolating operators for state creation and annihilation, constructed to have significant overlap onto the state of interest, this Euclidean time evolution, and thus exponential growth in noise, can be reduced. In this Letter, it is demonstrated for the 1+1-dimensional Schwinger model how one can construct such interpolating operators for classical LQFT calculations using small-scale quantum computation. Ultimately, the extension of this approach to the more complex theory of QCD, along with advancement in quantum hardware, could enable an significant acceleration of LQFT computations for nuclear physics.

{\bf The Schwinger model:} The Schwinger model~\cite{PhysRev.128.2425}, which describes the theory of quantum electrodynamics in one space and one time dimension, is a prototypical lattice gauge theory that shares a number of key features with QCD, including confinement. This model thus provides a simplified framework to test new algorithms and approaches to LQFT studies.
The theory describes fermions as a two-component spinor field $\psi$, with mass $m$, coupled via charge $g$ to an electromagnetic field $A_\mu$, with the Lagrangian defined as
\begin{align}
\label{eq:schwinger-lagrangian}
    \mathcal{L} &= \bar{\psi} \left(i \slashed{D} - m \right) \psi - \frac{1}{4} F_{\mu\nu} F^{\mu\nu},
\end{align}
where the covariant derivative $\slashed{D}$ and field strength tensor $F_{\mu\nu}$ are defined in terms of $A_\mu$.
A discretized formulation of the 1+1D Schwinger model can be defined on a staggered space-time lattice via the Kogut-Susskind prescription~\cite{PhysRevD.11.395}; the staggered fermion field $\phi(x_n)=\phi_n$ is defined with the two components of the fermion field occupying even and odd sites by
\begin{equation}
    \phi_n = \begin{cases}
    \sqrt{a} \psi_{e^-}(x_n), \text{  $n$ even,}\\
    \sqrt{a} \psi_{e^+}(x_n), \text{  $n$ odd}.
    \end{cases}
\end{equation}
In temporal gauge ($A_0=0$), the remaining spatial component of the gauge field is encoded on links connecting adjacent staggered sites $x_n$ and $x_{n+1}$ by
\begin{equation}
\label{eq:link-variables-1}
    \hat{U}(x_n, x_{n+1}) = e^{i a g \hat{A}_1(x_n)} \equiv  e^{i  \hat{\theta}_n}.
\end{equation}
Electric flux operators can be defined in terms of $\hat{A}_1$ as 
\begin{equation}
\label{eq:link-variables-2}
    \hat{\ell}_n = -\frac{1}{g}\frac{d\hat{A}_1}{dt}, 
\end{equation}
and can be interpreted as acting on the Fock space of links connecting sites: with $\ell_n$ denoting the value of the electric flux at the link connecting sites $n$ and $n+1$,
\begin{equation}
\label{eq:link-variables-4}
 \hat{\ell}_n \ket{\ell_n} = \ell_n \ket{\ell_n},\,\ell_n \in \mathbb{Z}\;\forall n.
\end{equation}
Ladder operators in this space can be defined as
\begin{equation}
\label{eq:link-variables-3}
    \hat{L}^{\pm}_{n} \equiv e^{\pm i \hat{\theta}_n}, \hspace{5mm} \hat{L}^{\pm}_{n} \ket{\ell_n} = \ket{\ell_n \pm 1}.
\end{equation}

Combining the link space with fermionic occupation numbers, a complete Fock space of states in this theory can be expressed as $\{ \big|\vec{n},\,\vec{\ell}\big\rangle \}$.
On a lattice with $N$ staggered sites ($N/2$ spatial sites), and with coupling constants $w = \frac{1}{2a}$ and $J = \frac{g^2 a}{2}$, the Hamiltonian of this theory can be expressed in terms of these operators as
\begin{align}
\label{eq:lattice-ham}\nonumber
    \hat{H}_{\mathrm{lat}} 
        =& iw \sum_{n=0}^{N-2} 
        (\hat{\phi}^\dagger_n \hat{L}^+_n \hat{\phi}_{n+1} - \mathrm{h.c.})
        \\\nonumber
        &- iw (\hat{\phi}^\dagger_{N-1} \hat{L}^+_{N-1} \hat{\phi}_{0} - \mathrm{h.c.}) \\
        &+ m \sum_{n=0}^{N-1} (-1)^n \hat{\phi}^\dagger_n \hat{\phi}_n
        + J \sum_{n=0}^{N-1} \hat{\ell}^2_n.
\end{align}
This theory has a simple spectrum of low-lying states of conserved parity quantum number; the first excited state is odd-parity, interpreted as the massive photon (the lightest vector meson), while the second excited state is the even-parity scalar $e^+e^-$ `meson'.

{\bf Classical computations of ground-state energies:} 
Using classical computation, energy levels of the Schwinger model can be obtained using standard Monte-Carlo (MC) methods. Here a local Hamiltonian MC method is studied~\cite{Hirsch1982MonteSystems}; details of the application of this approach to the 1+1D Schwinger model are given in Ref.~\cite{Schiller:1983sj}.
In this formalism, ground-state energy levels can be determined by the analysis of correlation functions $G(\tau )$, defined in terms of the expectation values of {\it interpolating operators} $\hat{\mathcal{O}}(x,\tau)$, which are constructed to create/annihilate states with the quantum numbers of a target state of interest at some Euclidean position $(x,\tau)$: 
\begin{equation}
\label{eq:two-point-corr-1}
    G(\tau ) =  
           \sum_x  \left[\left\langle 
                \hat{\mathcal{O}}(x,\tau)
                \hat{\mathcal{O}}^\dagger(0,0))
            \right\rangle
        -   \left\langle
                \hat{\mathcal{O}}(x,\tau)  
            \right\rangle
            \left\langle        
                \hat{\mathcal{O}}^\dagger(0,0)   
            \right\rangle\right].
\end{equation}
Here, a state is created at some initial spatial position and time $(x=0,t=0)$, and annihilated $\tau$ Euclidean time-steps later. The sum over $x$ projects onto the zero-momentum sector. 
This correlation depends on the energy gaps between the ground state (vacuum) of the system $\ket{\Omega}$ and the tower of excitations coupled to the vacuum through $\hat{\mathcal{O}}$:
\begin{equation}
G(\tau)=\sum_n\left|\langle n|\mathcal{O}|\Omega\rangle\right|^2 e^{-(E_n-E_\Omega)\tau}.
\end{equation}
Numerically, the energy gap to the lowest state of interest is determined from the asymptotic value of the {\it effective mass} function:
\begin{equation}
\label{eq:eff-mass}
\begin{aligned}
   M_{\mathrm{eff}}(\tau) = \frac{1}{a} 
                        \log\left(
                        \frac
                            {G(\tau)}
                            {G(\tau + a)}
                        \right) \underset{\tau \rightarrow \infty}{\rightarrow} (E_\mathcal{O}-E_\Omega).
\end{aligned}
\end{equation}

Interpolating operators for lattice field theories can be constructed by inspection, and often the simplest operators which have the quantum numbers of the state of interest are chosen. 
For the Schwinger model, the lowest-energy excitation is described by the lightest vector meson, $V^-$, a massive photon.
An interpolating operator for this state can be constructed by following the same prescription as for odd-parity meson states in staggered lattice formulations of QCD~\cite{Golterman1986StaggeredMesons}:
\begin{equation}
\begin{aligned}
\label{eq:m01}
    \hat{O}_V(x_n,\tau_j)
        &= \hat{\phi}^\dagger_{n,j} 
            \hat{L}^{+}_{n,j} \hat{\phi}_{n+1,j}
            - \hat{\phi}^\dagger_{n,j} 
            \hat{L}^{-}_{n-1,j} \hat{\phi}_{n-1,j}.
\end{aligned}
\end{equation}
Physically, the operator $\hat{O}_V$ creates an $e^+ e^-$ pair on sites $x_{n-1}$ and $x_{n}$, and a second $e^- e^+$ pair on sites $x_{n}$ and $x_{n+1}$. The relative minus sign between the two terms ensures that the construction has odd parity. 

Improved interpolating operators for LQFT calculations can be constructed classically via a variational approach: rather than a single interpolating operator, a set of operators with the same quantum numbers is chosen, and the resulting system is diagonalized via a generalised eigenvalue problem (GEVP) to achieve an optimized ground-state energy extraction in that sector~\cite{Michael:1985ne,Luscher:1990ck,Blossier2009OnTheory}. This approach has had tremendous success, particularly in the area of meson spectroscopy studies in LQFT~\cite{Wilson:2015dqa}.
Nevertheless, the classical variational method is computationally expensive, scaling quadratically with the size of the basis, and in particular using a large variational basis requires high-statistics numerical calculations to ensure a non-singular covariance matrix. 
This method thus remains intractable for many studies, such as state-of-the-art QCD calculations of nuclear systems~\cite{Joo2019StatusBeyond}. 
In this work, an alternative variational approach to interpolating operator construction is explored, in which the expense of a classical variational method is replaced with small-scale computations on quantum hardware.

{\bf Quantum approaches to Schwinger model dynamics:}
Several years ago, the first experimental demonstration of a digital quantum simulation of a lattice gauge theory was achieved by realizing the Schwinger model on a few-qubit trapped-ion quantum device~\cite{Hauke2013QuantumSimulation, Muschik2017U1Simulators, Martinez2016Real-timeComputer,Kokail:2018eiw}. 
Recently, the two-spatial-site Schwinger model has also been studied on IBM's superconducting quantum hardware\footnote{These quantum calculations use an equivalent formulation of the Schwinger model based on bosonic degrees of freedom, related by a Jordan-Wigner transformation~\cite{Ortiz2001QuantumAlgorithms} to the formulation described here.}~\cite{Klco:2018kyo}. 
In that work, the ground state energy level of the theory in was extracted using the variational quantum eigensolver (VQE) method~\cite{Peruzzo2014}. In the VQE approach, a sequence of unitary operators $U^{(i)}(\vec{\theta})$, implemented as a sequence of one- and two-qubit gates, are tuned using variational parameters $\theta$ to transform an initial, easy-to-prepare state $\ket{0}$ into an approximation of the ground state of the system in a given symmetry sector, $\ket{G}$:
\begin{align}
    \ket{G} \approx \ket{G,\vec\theta} = U^{(m)}(\vec{\theta}_m) U^{(m-1)}(\vec{\theta}_{m-1}) \ldots  U^{(1)}(\vec{\theta}_1) \ket{0}.\label{eq:vqs}
\end{align}
From this construction, an approximate value of the ground-state energy of the system can be calculated.
In the study of Klco et~al.~\cite{Klco:2018kyo}, which presents a formulation of the Schwinger model on quantum hardware which has a natural relation to classical approaches to the theory, explicit electric flux degrees of freedom are retained in the basis of states studied using the VQE approach. To render the Schwinger model Hilbert space in this description finite thus requires truncating the possible values of flux on each link.
This can be achieved by enforcing 
\begin{align}
\label{eq:trunc-1}
        \ell_i^2 \leq \Lambda^2, \,\forall\;i = 0,\ldots,N-1, \hspace{3mm} \sum_{i=0}^{N-1} \ell_i^2 \leq \tilde{\Lambda}^2,
\end{align}
for some choice of truncation $\{\Lambda^2,\tilde{\Lambda}^2\}$; harsher truncations result in larger systematic uncertainties in the determined energy level, but require fewer qubits for simulation. Naturally, the small system sizes accommodated by NISQ-era quantum-computing hardware result in additional finite-size systematic uncertainties in numerical studies.

Given the present status of quantum computation, scaling these studies to determine ground-state energies of systems of physical interest is a long-term challenge. Using VQE calculations in hybrid approaches to accelerate classical LQFT computations, which can be more easily scaled at the present time, however, offers the potential of near-term exploitation of these new tools.

{\bf Quantum-improved interpolating operators:}
Here, it is proposed to use the information encoded in a variationally-obtained approximation to a ground-state wavefunction to construct an improved interpolating operator for use in classical LQFT computations of that state. 
This is achieved via a two-step approximation process: first, VQE calculations are used to yield approximate representations of the wavefunctions of both the dynamical vacuum and the ground state of the symmetry sector of interest; second, a linear combination of operators is optimized to maximize the transition matrix element between the vacuum and the state of interest.
This procedure can be considered as analogous to the classical variational approach to interpolating operator construction, and may have significant advantages in a near-future era of efficient small-scale quantum computation. In particular, as is demonstrated here, one can undertake a quantum computation in a truncated Hilbert space on a small lattice volume, and use this to determine a significantly reduced basis of operators to compute classically on the full state space.

As an explicit example, for the Schwinger model one might aim to study the first excited state of the theory (the odd-parity massive photon). As outlined in Eq.~\eqref{eq:vqs}, VQE computations can provide approximate descriptions of both this state, $|{V^-,\vec\theta'}\rangle$, and the vacuum of the theory $|{\Omega,\vec\theta}\rangle$.
Acting with a set of interpolating operators $\{\hat{O}_k\}$ on the ground state provides a variational basis that can be used to approximate the massive photon:
\begin{align}
    |\widetilde{V}^-\rangle=\sum_k\alpha_k\hat{O}_k|\Omega,\vec\theta\rangle.
\end{align}
The operators $\hat{O}_k$ can be defined, for example, in terms of field and link operators $\hat{\phi}_n$, $\hat{\ell}_n$, $\hat{L}_n^{\pm}$, and should be constructed to transition between the vacuum and the symmetry sector of interest.
With an estimate for the matrix elements
$\langle V^-,\vec\theta'|\hat{O}_k|\Omega,\vec\theta\rangle$, a classical computer can be used to optimize the overlap 
$|\langle V^-,\vec\theta|\widetilde{V}^-\rangle|$ (with appropriate normalization) with respect to $\vec\alpha$.
This optimization defines an improved interpolating operator 
\begin{align}
    \hat O_{\rm VQE}=\sum_k\alpha_k\hat O_k,\label{eq:OVQE}
\end{align}
which has suppressed overlap onto excited states and can be implemented in a classical Euclidean MC calculation.
Ultimately, the matrix elements $\langle V^-,\vec\theta'|\hat{O}_k|\Omega,\vec\theta\rangle$ needed for this approach will be evaluated on quantum hardware. Within the present formalism, the Hamiltonian has been block-diagonalized into distinct symmetry sectors, whereas the operators of interest transition between these sectors. Nevertheless, in a space concatenating the two sectors, the operators can be written as products of Pauli operators such that projective measurements can be made in an approach similar to that in Refs.~\cite{Reiner2019,Kokail:2018eiw,Peruzzo2014,Ortiz2001QuantumAlgorithms}.
In the near term, these matrix elements can also be calculated directly on a classical computer given the quantum gate definitions and the angles \{$\vec\theta$,$\vec{\theta}'$\} which encode the approximate wavefunctions; for example, given the 2-layer depth of the variational circuits implemented for the Schwinger model on NISQ devices in Ref.~\cite{Klco:2018kyo}, this can be undertaken in $\mathcal{O}(N^2)$ operations for an $N$-dimensional VQE basis, which is still computationally tractable on intermediate system sizes where exact diagonalization, at $\mathcal{O}(N^3)$, is not.

{\bf Implementation for the 1+1D Schwinger model:}
The proposed approach to interpolating operator construction is implemented in the context of the 1+1D Schwinger model defined by Eq.~\eqref{eq:schwinger-lagrangian}.
The parameters of the Hamiltonian are chosen to match those in recent studies of this model on quantum devices~\cite{Klco:2018kyo}: $J/\omega=5/3$ and $m/\omega=1/6$, and the temporal lattice spacing is set as $\Delta t = a/5$ based on previous Monte-Carlo studies~\cite{Schiller:1983sj}. 

\begin{figure}[!t]
\includegraphics[width=\columnwidth]{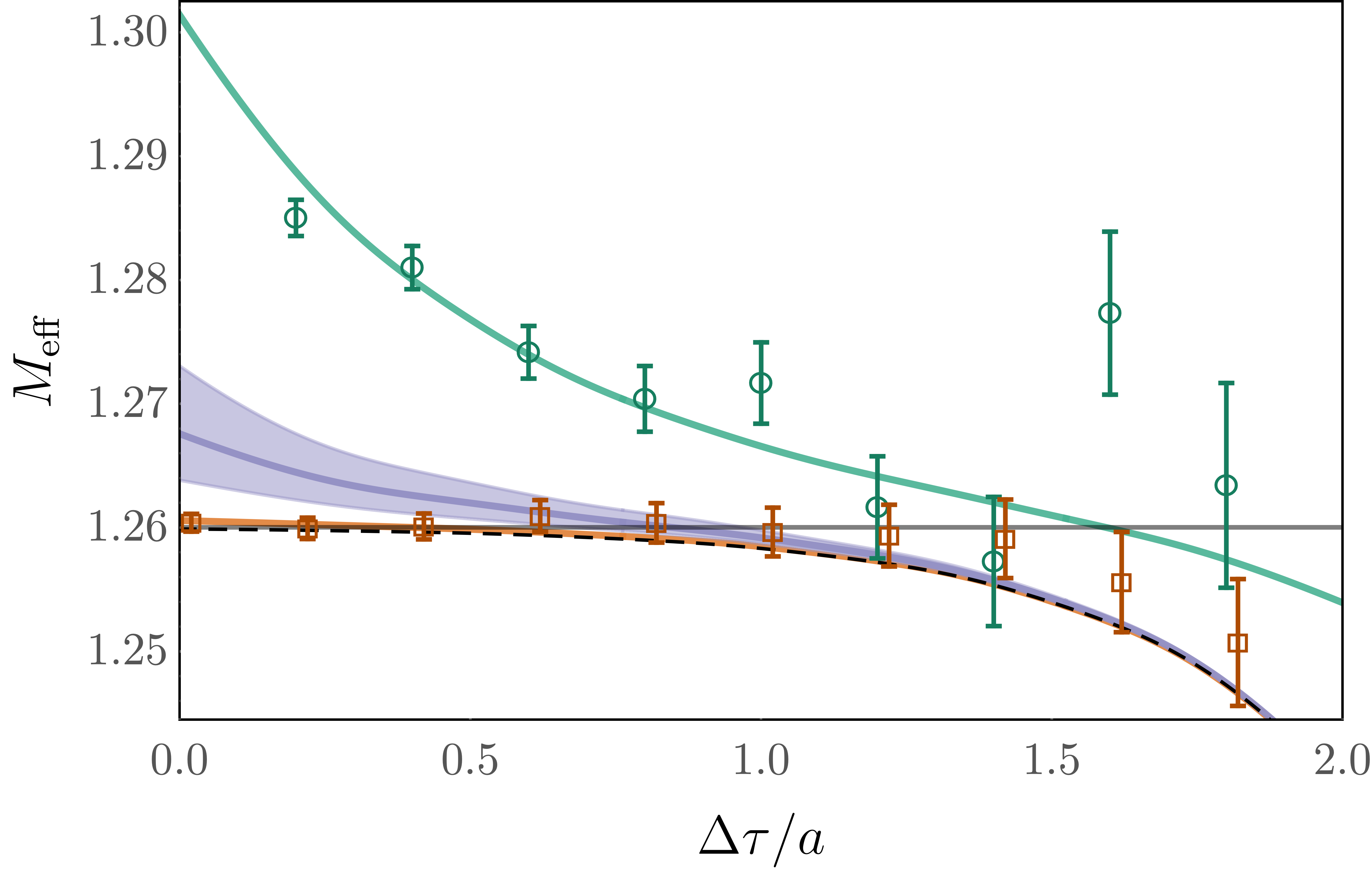}
\caption{\label{fig:effmass}
	Effective mass functions (Eq.~\eqref{eq:eff-mass}) constructed exactly via diagonalization (solid lines), and obtained from MC computations (open points), for the lightest vector meson. The green circles and red squares denote results obtained using the benchmark (Eq.~\eqref{eq:m01}) and quantum-improved (Eq.~\eqref{eq:OVQE}) operators, respectively. The black dashed line shows the result obtained solving an exact form of the classical GEVP using the same basis of interpolating operators, while the purple solid band shows the exact result for the optimized operator, given some uncertainty on the VQE input into its construction as described in the text.
	The large time behaviour is a consequence of the backward propagating states around the boundary at finite Euclidean time.}
\end{figure}

For direct comparison, the mass of the lightest vector meson in this model can be computed exactly and extracted using the local Hamiltonian MC method~\cite{Hirsch1982MonteSystems,Schiller:1983sj}. Here, a system with 4 spatial sites (8 staggered sites), and 40 temporal (80 staggered sites) is studied, with the standard interpolating operator defined in Eq.~\eqref{eq:m01} used as a benchmark; the effective mass from a numerical computation with $10^7$ configurations sampled  is shown in Fig.~\ref{fig:effmass}.
The effective mass with this operator is also reconstructed exactly from the numerical diagonalization of Eq.~\eqref{eq:lattice-ham}.
This diagonalization is performed with respect to a truncation of the electric fluxes, $\{\Lambda^2,\tilde{\Lambda}^2\}=\{1,8\}$, that is chosen to encode the same physics that is being sampled within the Euclidean MC setup.

A quantum-improved interpolating operator, constructed via the approach proposed here, can also be investigated. 
First, the even and odd-parity ground-states are obtained from the corresponding exact solution, which is used as a proxy for a VQE in this proof-of-principle demonstration. 
A linear combination of up to 6 operators defined in terms of electric flux operators $\hat{\ell}$ (Eq.~\eqref{eq:link-variables-4}), detailed in the supplementary material, is then optimized to maximise the overlap with the target state.
This improved operator, of the form Eq.~\eqref{eq:OVQE}, is then used to construct a correlator (Eq.~\eqref{eq:two-point-corr-1}) which has significantly improved overlap with the lowest-lying negative-parity state.
Figure \ref{fig:effmass} displays the corresponding effective masses obtained for both the Monte Carlo ensemble and those constructed from exact diagonalisation.
The effective mass obtained from a MC computation with the quantum-improved interpolating operator leads to a far better constrained mass extraction at the same statistics than the benchmark operator.
Moreover, the quantum-improved operator produces an exact effective mass curve which is indistinguishable, on the scale of Fig.~\ref{fig:effmass}, from that obtained via an exact version of the classical generalized eigenvalue program.

Naturally, in a true quantum computation the VQE wavefunctions will be only approximately determined, with statistical uncertainties on the variational parameters \{$\vec\theta$,$\vec{\theta}'$\}, and the corresponding matrix elements $\langle V^-,\vec\theta'|\hat O_k|\Omega,\vec\theta\rangle$ will be similarly limited in measurement precision.
The effect of such uncertainties on the definition of the quantum-improved interpolating operator $\hat O_{\rm VQE}$ is investigated by taking a 0.05 radian error on the variational parameters, corresponding to approximately 15\% error on the coefficients of expansion. These uncertainties are roughly equivalent to the fidelities obtained in recent studies of the Schwinger model using modern quantum hardware~\cite{Kokail:2018eiw}. Fig.~\ref{fig:effmass} shows the corresponding uncertainty in the VQE-improved correlation function.
Importantly, even with an imperfect quantum computation, the quantum-improved operator offers improved isolation of the ground state in comparison with the benchmark operator.

As described, it is evident that a quantum calculation has the potential to improve the results of a conventional MC computation.
Nevertheless, the scaling of quantum hardware to encode larger Hilbert spaces is expected to be an enduring problem.
It would therefore be of tremendous value if quantum simulations in a (significantly) truncated Hilbert space can still produce improved operators for a Euclidean MC calculation---which can more readily be scaled to larger systems.
Restricting the gauge space, by truncating the gauge link variables, and limiting the spatial volume, are two natural ways to reduce the Hilbert space for simulations on quantum hardware.

\begin{figure}[!t]
\includegraphics[width=\columnwidth]{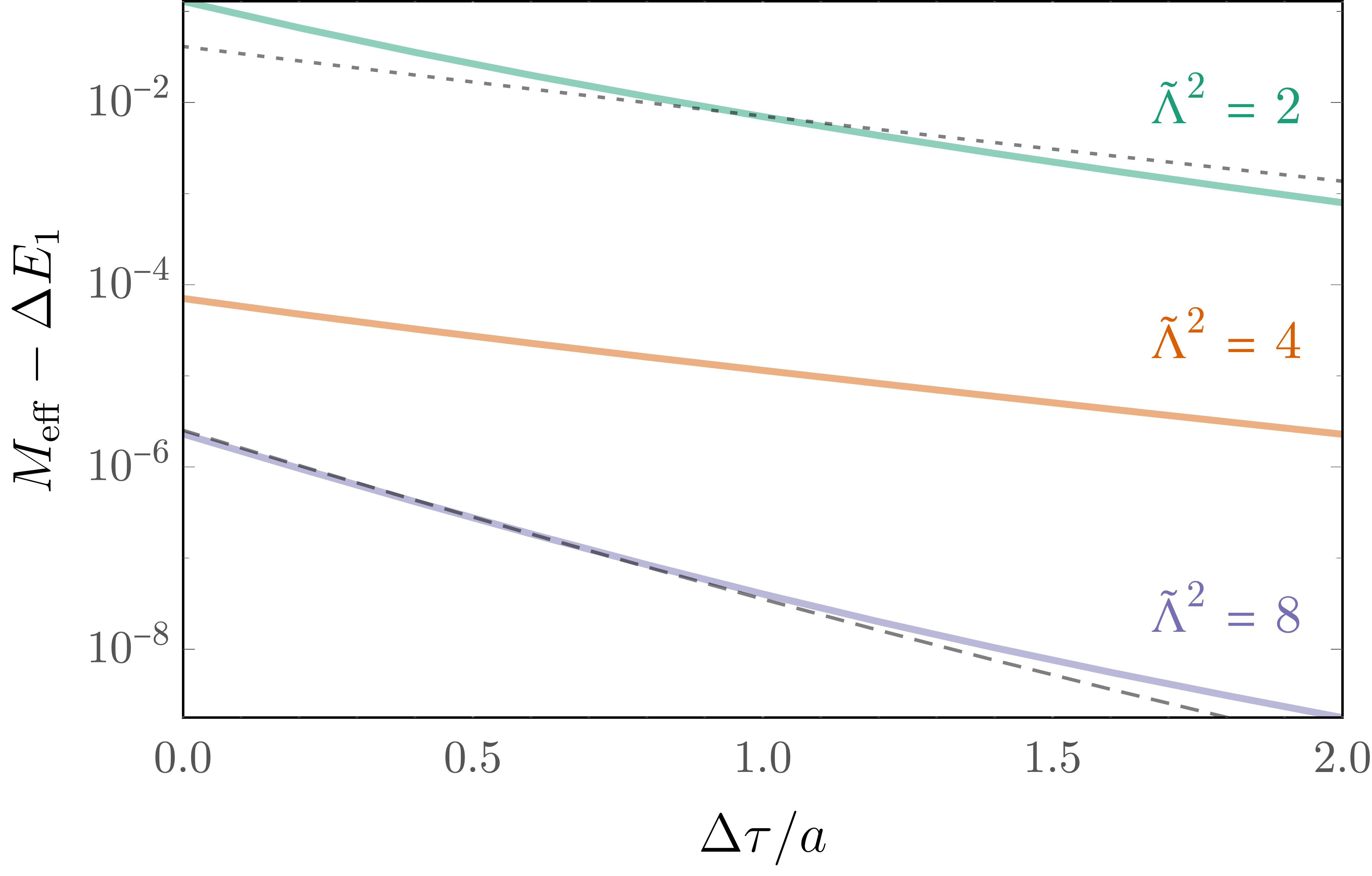}
\caption{Exact effective mass functions in an $L=6$ spatial-site Schwinger model computation, calculated using quantum-improved interpolating operators constructed on smaller Hilbert spaces. Green, orange, and purple curves (moving vertically down the figure) show results obtained via optimization of systems with flux truncations $\tilde{\Lambda}^2$ = \{2,4,8\} on an $L=4$ system, respectively. The grey dashed and dotted curves indicate the exact classical GEVP solved on the full Hilbert space, and the exact result corresponding to the native interpolating operator (Eq.~\eqref{eq:m01}) respectively.
\label{fig:truncate}}
\end{figure}

Figure \ref{fig:truncate} displays an exact calculation of effective masses, for different operator constructions, for a Schwinger model with 6 spatial sites.
Each contour shows results obtained using quantum-improved operators of the form Eq.~\eqref{eq:OVQE} optimized on smaller Hilbert spaces than the space they are applied to. In particular, optimizations are undertaken on the 4 spatial-site system, and with different levels of truncation on the total link-square parameter $\tilde\Lambda^2$.
It is clear that even on this smaller system, all operator constructions with truncations $\tilde{\Lambda}^2>2$ perform significantly better than the naive operator presented in Fig.~\ref{fig:effmass}.
For comparison, we note that the dimensionality of the truncated Hilbert space with $L=4$ and $\tilde{\Lambda}^2=4$ is 10 in the zero-momentum, odd-parity subspace, compared to 100 for the same subspace of the $\tilde{\Lambda}^2=12$ $L=6$ system.
It should also be noted that the improved operators from the VQE on truncated spaces approach the effective mass of GEVP computed on the exact system for the same operator basis.

The potential advantages of this approach are clear.
In particular, the scaling of quantum simulations is challenging, and the approach presented here exploits the strengths of both quantum and classical computation; an imprecise interpolating operator extracted from a noisy quantum simulation can still outperform a more naive interpolating operator in a classical calculation, while the classical computation using that operator can be much more easily scaled to large space-time volumes than the quantum calculation.

{\bf Discussion:}
Here it is demonstrated how improved interpolating operators for lattice field theory calculations can be constructed using information from VQE computations on NISQ-era quantum hardware. 
A key feature is of this method is that it does not require the large quantum systems that would be needed for a direct calculation of the physics of interest on the quantum machine, but can still accelerate the classical calculation analogously to a classical variational approach, with a potentially-significant reduction in classical computing resources.
Moreover, while complicated properties of a ground-state system are challenging to access using quantum hardware and have not yet been directly computed on quantum hardware even for the simple Schwinger model system, an optimal interpolating operator obtained via the procedure proposed here can be used in classical computations to accelerate the evaluation of many other properties of the state, including its interactions with external probes.

Extensions of this approach to field theories of phenomenological interest such as QCD could proceed via methods similar to those investigated in Refs.~\cite{Brower:1997ha, Wiese:2014rla}. Ultimately, there is tremendous opportunity for NISQ-era quantum devices to improve classical field theory calculations. The approach described here represents a clear step towards realizing this goal.

\bibliography{refs.bib}

\begin{thebibliography}{26}
\expandafter\ifx\csname natexlab\endcsname\relax\def\natexlab#1{#1}\fi
\expandafter\ifx\csname bibnamefont\endcsname\relax
  \def\bibnamefont#1{#1}\fi
\expandafter\ifx\csname bibfnamefont\endcsname\relax
  \def\bibfnamefont#1{#1}\fi
\expandafter\ifx\csname citenamefont\endcsname\relax
  \def\citenamefont#1{#1}\fi
\expandafter\ifx\csname url\endcsname\relax
  \def\url#1{\texttt{#1}}\fi
\expandafter\ifx\csname urlprefix\endcsname\relax\def\urlprefix{URL }\fi
\providecommand{\bibinfo}[2]{#2}
\providecommand{\eprint}[2][]{\url{#2}}

\bibitem[{\citenamefont{Wilson}(1974)}]{Wilson:1974sk}
\bibinfo{author}{\bibfnamefont{K.~G.} \bibnamefont{Wilson}},
  \bibinfo{journal}{Phys. Rev.} \textbf{\bibinfo{volume}{D10}},
  \bibinfo{pages}{2445} (\bibinfo{year}{1974}).

\bibitem[{\citenamefont{Gattringer and Lang}(2010)}]{Gattringer:2010zz}
\bibinfo{author}{\bibfnamefont{C.}~\bibnamefont{Gattringer}} \bibnamefont{and}
  \bibinfo{author}{\bibfnamefont{C.~B.} \bibnamefont{Lang}},
  \bibinfo{journal}{Lect. Notes Phys.} \textbf{\bibinfo{volume}{788}},
  \bibinfo{pages}{1} (\bibinfo{year}{2010}).

\bibitem[{\citenamefont{Aoki et~al.}(2019)}]{Aoki:2019cca}
\bibinfo{author}{\bibfnamefont{S.}~\bibnamefont{Aoki}} \bibnamefont{et~al.}
  (\bibinfo{collaboration}{Flavour Lattice Averaging Group})
  (\bibinfo{year}{2019}), \eprint{1902.08191}.

\bibitem[{\citenamefont{Lehner et~al.}(2019)}]{Lehner:2019wvv}
\bibinfo{author}{\bibfnamefont{C.}~\bibnamefont{Lehner}} \bibnamefont{et~al.}
  (\bibinfo{collaboration}{USQCD}) (\bibinfo{year}{2019}), \eprint{1904.09479}.

\bibitem[{\citenamefont{Jo{\'{o}} et~al.}(2019)\citenamefont{Jo{\'{o}}, Jung,
  Christ, Detmold, Edwards, Savage, and Shanahan}}]{Joo2019StatusBeyond}
\bibinfo{author}{\bibfnamefont{B.}~\bibnamefont{Jo{\'{o}}}},
  \bibinfo{author}{\bibfnamefont{C.}~\bibnamefont{Jung}},
  \bibinfo{author}{\bibfnamefont{N.~H.} \bibnamefont{Christ}},
  \bibinfo{author}{\bibfnamefont{W.}~\bibnamefont{Detmold}},
  \bibinfo{author}{\bibfnamefont{R.}~\bibnamefont{Edwards}},
  \bibinfo{author}{\bibfnamefont{M.}~\bibnamefont{Savage}}, \bibnamefont{and}
  \bibinfo{author}{\bibfnamefont{P.}~\bibnamefont{Shanahan}}
  (\bibinfo{year}{2019}), \urlprefix\url{http://arxiv.org/abs/1904.09725}.

\bibitem[{\citenamefont{Carlson et~al.}(2018)\citenamefont{Carlson, Dean,
  Kaplan, Preskill, Roche, Savage, and Troyer}}]{CarlsonQuantumPhysics}
\bibinfo{author}{\bibfnamefont{J.}~\bibnamefont{Carlson}},
  \bibinfo{author}{\bibfnamefont{D.~J.} \bibnamefont{Dean}},
  \bibinfo{author}{\bibfnamefont{D.}~\bibnamefont{Kaplan}},
  \bibinfo{author}{\bibfnamefont{J.}~\bibnamefont{Preskill}},
  \bibinfo{author}{\bibfnamefont{K.}~\bibnamefont{Roche}},
  \bibinfo{author}{\bibfnamefont{M.~J.} \bibnamefont{Savage}},
  \bibnamefont{and} \bibinfo{author}{\bibfnamefont{M.}~\bibnamefont{Troyer}},
  \bibinfo{type}{Tech. Rep.} (\bibinfo{year}{2018}),
  \urlprefix\url{http://www.int.washington.edu/PROGRAMS/17-66W/QuantumComputing_NUCLEARPHYSICS_FINAL_pdf.pdf}.

\bibitem[{\citenamefont{Preskill}(2018)}]{Preskill2018}
\bibinfo{author}{\bibfnamefont{J.}~\bibnamefont{Preskill}},
  \bibinfo{journal}{{Quantum}} \textbf{\bibinfo{volume}{2}},
  \bibinfo{pages}{79} (\bibinfo{year}{2018}), ISSN \bibinfo{issn}{2521-327X},
  \urlprefix\url{https://doi.org/10.22331/q-2018-08-06-79}.

\bibitem[{\citenamefont{Schwinger}(1962)}]{PhysRev.128.2425}
\bibinfo{author}{\bibfnamefont{J.}~\bibnamefont{Schwinger}},
  \bibinfo{journal}{Phys. Rev.} \textbf{\bibinfo{volume}{128}},
  \bibinfo{pages}{2425} (\bibinfo{year}{1962}),
  \urlprefix\url{https://link.aps.org/doi/10.1103/PhysRev.128.2425}.

\bibitem[{\citenamefont{Kogut and Susskind}(1975)}]{PhysRevD.11.395}
\bibinfo{author}{\bibfnamefont{J.}~\bibnamefont{Kogut}} \bibnamefont{and}
  \bibinfo{author}{\bibfnamefont{L.}~\bibnamefont{Susskind}},
  \bibinfo{journal}{Phys. Rev. D} \textbf{\bibinfo{volume}{11}},
  \bibinfo{pages}{395} (\bibinfo{year}{1975}),
  \urlprefix\url{https://link.aps.org/doi/10.1103/PhysRevD.11.395}.

\bibitem[{\citenamefont{Hirsch et~al.}(1982)\citenamefont{Hirsch, Sugar,
  Scalapino, and Blankenbecler}}]{Hirsch1982MonteSystems}
\bibinfo{author}{\bibfnamefont{J.~E.} \bibnamefont{Hirsch}},
  \bibinfo{author}{\bibfnamefont{R.~L.} \bibnamefont{Sugar}},
  \bibinfo{author}{\bibfnamefont{D.~J.} \bibnamefont{Scalapino}},
  \bibnamefont{and}
  \bibinfo{author}{\bibfnamefont{R.}~\bibnamefont{Blankenbecler}},
  \bibinfo{journal}{Physical Review B} \textbf{\bibinfo{volume}{26}},
  \bibinfo{pages}{5033} (\bibinfo{year}{1982}), ISSN \bibinfo{issn}{01631829}.

\bibitem[{\citenamefont{Schiller and Ranft}(1983)}]{Schiller:1983sj}
\bibinfo{author}{\bibfnamefont{A.}~\bibnamefont{Schiller}} \bibnamefont{and}
  \bibinfo{author}{\bibfnamefont{J.}~\bibnamefont{Ranft}},
  \bibinfo{journal}{Nucl. Phys.} \textbf{\bibinfo{volume}{B225}},
  \bibinfo{pages}{204} (\bibinfo{year}{1983}).

\bibitem[{\citenamefont{Golterman}(1986)}]{Golterman1986StaggeredMesons}
\bibinfo{author}{\bibfnamefont{M.~F.~L.} \bibnamefont{Golterman}},
  \bibinfo{journal}{Nuclear Physics, Section B} \textbf{\bibinfo{volume}{273}},
  \bibinfo{pages}{663} (\bibinfo{year}{1986}), ISSN \bibinfo{issn}{05503213}.

\bibitem[{\citenamefont{Michael}(1985)}]{Michael:1985ne}
\bibinfo{author}{\bibfnamefont{C.}~\bibnamefont{Michael}},
  \bibinfo{journal}{Nucl. Phys.} \textbf{\bibinfo{volume}{B259}},
  \bibinfo{pages}{58} (\bibinfo{year}{1985}).

\bibitem[{\citenamefont{Luscher and Wolff}(1990)}]{Luscher:1990ck}
\bibinfo{author}{\bibfnamefont{M.}~\bibnamefont{Luscher}} \bibnamefont{and}
  \bibinfo{author}{\bibfnamefont{U.}~\bibnamefont{Wolff}},
  \bibinfo{journal}{Nucl. Phys.} \textbf{\bibinfo{volume}{B339}},
  \bibinfo{pages}{222} (\bibinfo{year}{1990}).

\bibitem[{\citenamefont{Blossier et~al.}(2009)\citenamefont{Blossier, Morte,
  Von~Hippel, Mendes, and Sommer}}]{Blossier2009OnTheory}
\bibinfo{author}{\bibfnamefont{B.}~\bibnamefont{Blossier}},
  \bibinfo{author}{\bibfnamefont{M.~D.} \bibnamefont{Morte}},
  \bibinfo{author}{\bibfnamefont{G.}~\bibnamefont{Von~Hippel}},
  \bibinfo{author}{\bibfnamefont{T.}~\bibnamefont{Mendes}}, \bibnamefont{and}
  \bibinfo{author}{\bibfnamefont{R.}~\bibnamefont{Sommer}},
  \bibinfo{journal}{Journal of High Energy Physics}  (\bibinfo{year}{2009}),
  \urlprefix\url{https://arxiv.org/pdf/0902.1265.pdf}.

\bibitem[{\citenamefont{Wilson et~al.}(2015)\citenamefont{Wilson, Brice\~no,
  Dudek, Edwards, and Thomas}}]{Wilson:2015dqa}
\bibinfo{author}{\bibfnamefont{D.~J.} \bibnamefont{Wilson}},
  \bibinfo{author}{\bibfnamefont{R.~A.} \bibnamefont{Brice\~no}},
  \bibinfo{author}{\bibfnamefont{J.~J.} \bibnamefont{Dudek}},
  \bibinfo{author}{\bibfnamefont{R.~G.} \bibnamefont{Edwards}},
  \bibnamefont{and} \bibinfo{author}{\bibfnamefont{C.~E.}
  \bibnamefont{Thomas}}, \bibinfo{journal}{Phys. Rev.}
  \textbf{\bibinfo{volume}{D92}}, \bibinfo{pages}{094502}
  (\bibinfo{year}{2015}), \eprint{1507.02599}.

\bibitem[{\citenamefont{Hauke et~al.}(2013)\citenamefont{Hauke, Marcos,
  Dalmonte, and Zoller}}]{Hauke2013QuantumSimulation}
\bibinfo{author}{\bibfnamefont{P.}~\bibnamefont{Hauke}},
  \bibinfo{author}{\bibfnamefont{D.}~\bibnamefont{Marcos}},
  \bibinfo{author}{\bibfnamefont{M.}~\bibnamefont{Dalmonte}}, \bibnamefont{and}
  \bibinfo{author}{\bibfnamefont{P.}~\bibnamefont{Zoller}},
  \bibinfo{journal}{Phys. Rev. X} \textbf{\bibinfo{volume}{3}},
  \bibinfo{pages}{041018} (\bibinfo{year}{2013}),
  \urlprefix\url{https://link.aps.org/doi/10.1103/PhysRevX.3.041018}.

\bibitem[{\citenamefont{Muschik et~al.}(2017)\citenamefont{Muschik, Heyl,
  Martinez, Monz, Schindler, Vogell, Dalmonte, Hauke, Blatt, and
  Zoller}}]{Muschik2017U1Simulators}
\bibinfo{author}{\bibfnamefont{C.}~\bibnamefont{Muschik}},
  \bibinfo{author}{\bibfnamefont{M.}~\bibnamefont{Heyl}},
  \bibinfo{author}{\bibfnamefont{E.}~\bibnamefont{Martinez}},
  \bibinfo{author}{\bibfnamefont{T.}~\bibnamefont{Monz}},
  \bibinfo{author}{\bibfnamefont{P.}~\bibnamefont{Schindler}},
  \bibinfo{author}{\bibfnamefont{B.}~\bibnamefont{Vogell}},
  \bibinfo{author}{\bibfnamefont{M.}~\bibnamefont{Dalmonte}},
  \bibinfo{author}{\bibfnamefont{P.}~\bibnamefont{Hauke}},
  \bibinfo{author}{\bibfnamefont{R.}~\bibnamefont{Blatt}}, \bibnamefont{and}
  \bibinfo{author}{\bibfnamefont{P.}~\bibnamefont{Zoller}},
  \bibinfo{journal}{New Journal of Physics} \textbf{\bibinfo{volume}{19}}
  (\bibinfo{year}{2017}), ISSN \bibinfo{issn}{13672630}.

\bibitem[{\citenamefont{Martinez et~al.}(2016)\citenamefont{Martinez, Muschik,
  Schindler, Nigg, Erhard, Heyl, Hauke, Dalmonte, Monz, Zoller
  et~al.}}]{Martinez2016Real-timeComputer}
\bibinfo{author}{\bibfnamefont{E.~A.} \bibnamefont{Martinez}},
  \bibinfo{author}{\bibfnamefont{C.~A.} \bibnamefont{Muschik}},
  \bibinfo{author}{\bibfnamefont{P.}~\bibnamefont{Schindler}},
  \bibinfo{author}{\bibfnamefont{D.}~\bibnamefont{Nigg}},
  \bibinfo{author}{\bibfnamefont{A.}~\bibnamefont{Erhard}},
  \bibinfo{author}{\bibfnamefont{M.}~\bibnamefont{Heyl}},
  \bibinfo{author}{\bibfnamefont{P.}~\bibnamefont{Hauke}},
  \bibinfo{author}{\bibfnamefont{M.}~\bibnamefont{Dalmonte}},
  \bibinfo{author}{\bibfnamefont{T.}~\bibnamefont{Monz}},
  \bibinfo{author}{\bibfnamefont{P.}~\bibnamefont{Zoller}},
  \bibnamefont{et~al.}, \bibinfo{journal}{Nature}
  \textbf{\bibinfo{volume}{534}}, \bibinfo{pages}{516} (\bibinfo{year}{2016}),
  ISSN \bibinfo{issn}{14764687}.

\bibitem[{\citenamefont{Kokail et~al.}(2019)}]{Kokail:2018eiw}
\bibinfo{author}{\bibfnamefont{C.}~\bibnamefont{Kokail}} \bibnamefont{et~al.},
  \bibinfo{journal}{Nature} \textbf{\bibinfo{volume}{569}},
  \bibinfo{pages}{355} (\bibinfo{year}{2019}), \eprint{1810.03421}.

\bibitem[{\citenamefont{Ortiz et~al.}(2001)\citenamefont{Ortiz, Gubernatis,
  Knill, and Laflamme}}]{Ortiz2001QuantumAlgorithms}
\bibinfo{author}{\bibfnamefont{G.}~\bibnamefont{Ortiz}},
  \bibinfo{author}{\bibfnamefont{J.~E.} \bibnamefont{Gubernatis}},
  \bibinfo{author}{\bibfnamefont{E.}~\bibnamefont{Knill}}, \bibnamefont{and}
  \bibinfo{author}{\bibfnamefont{R.}~\bibnamefont{Laflamme}},
  \bibinfo{journal}{Phys. Rev. A} \textbf{\bibinfo{volume}{64}},
  \bibinfo{pages}{022319} (\bibinfo{year}{2001}),
  \urlprefix\url{https://link.aps.org/doi/10.1103/PhysRevA.64.022319}.

\bibitem[{\citenamefont{Klco et~al.}(2018)\citenamefont{Klco, Dumitrescu,
  McCaskey, Morris, Pooser, Sanz, Solano, Lougovski, and
  Savage}}]{Klco:2018kyo}
\bibinfo{author}{\bibfnamefont{N.}~\bibnamefont{Klco}},
  \bibinfo{author}{\bibfnamefont{E.~F.} \bibnamefont{Dumitrescu}},
  \bibinfo{author}{\bibfnamefont{A.~J.} \bibnamefont{McCaskey}},
  \bibinfo{author}{\bibfnamefont{T.~D.} \bibnamefont{Morris}},
  \bibinfo{author}{\bibfnamefont{R.~C.} \bibnamefont{Pooser}},
  \bibinfo{author}{\bibfnamefont{M.}~\bibnamefont{Sanz}},
  \bibinfo{author}{\bibfnamefont{E.}~\bibnamefont{Solano}},
  \bibinfo{author}{\bibfnamefont{P.}~\bibnamefont{Lougovski}},
  \bibnamefont{and} \bibinfo{author}{\bibfnamefont{M.~J.}
  \bibnamefont{Savage}}, \bibinfo{journal}{Phys. Rev.}
  \textbf{\bibinfo{volume}{A98}}, \bibinfo{pages}{032331}
  (\bibinfo{year}{2018}), \eprint{1803.03326}.

\bibitem[{\citenamefont{Peruzzo et~al.}(2014)\citenamefont{Peruzzo, McClean,
  Shadbolt, Yung, Zhou, Love, Aspuru-Guzik, and O'Brien}}]{Peruzzo2014}
\bibinfo{author}{\bibfnamefont{A.}~\bibnamefont{Peruzzo}},
  \bibinfo{author}{\bibfnamefont{J.}~\bibnamefont{McClean}},
  \bibinfo{author}{\bibfnamefont{P.}~\bibnamefont{Shadbolt}},
  \bibinfo{author}{\bibfnamefont{M.-H.} \bibnamefont{Yung}},
  \bibinfo{author}{\bibfnamefont{X.-Q.} \bibnamefont{Zhou}},
  \bibinfo{author}{\bibfnamefont{P.~J.} \bibnamefont{Love}},
  \bibinfo{author}{\bibfnamefont{A.}~\bibnamefont{Aspuru-Guzik}},
  \bibnamefont{and} \bibinfo{author}{\bibfnamefont{J.~L.}
  \bibnamefont{O'Brien}}, \bibinfo{journal}{Nature Communications}
  \textbf{\bibinfo{volume}{5}}, \bibinfo{pages}{4213 EP }
  (\bibinfo{year}{2014}), \bibinfo{note}{article},
  \urlprefix\url{https://doi.org/10.1038/ncomms5213}.

\bibitem[{\citenamefont{Reiner et~al.}(2019)\citenamefont{Reiner,
  Wilhelm-Mauch, SchÃ¶n, and Marthaler}}]{Reiner2019}
\bibinfo{author}{\bibfnamefont{J.-M.} \bibnamefont{Reiner}},
  \bibinfo{author}{\bibfnamefont{F.}~\bibnamefont{Wilhelm-Mauch}},
  \bibinfo{author}{\bibfnamefont{G.}~\bibnamefont{SchÃ¶n}}, \bibnamefont{and}
  \bibinfo{author}{\bibfnamefont{M.}~\bibnamefont{Marthaler}},
  \bibinfo{journal}{Quantum Science and Technology}
  \textbf{\bibinfo{volume}{4}}, \bibinfo{pages}{035005} (\bibinfo{year}{2019}),
  \urlprefix\url{https://doi.org/10.1088%2F2058-9565%2Fab1e85}.

\bibitem[{\citenamefont{Brower et~al.}(1999)\citenamefont{Brower,
  Chandrasekharan, and Wiese}}]{Brower:1997ha}
\bibinfo{author}{\bibfnamefont{R.}~\bibnamefont{Brower}},
  \bibinfo{author}{\bibfnamefont{S.}~\bibnamefont{Chandrasekharan}},
  \bibnamefont{and} \bibinfo{author}{\bibfnamefont{U.~J.} \bibnamefont{Wiese}},
  \bibinfo{journal}{Phys. Rev.} \textbf{\bibinfo{volume}{D60}},
  \bibinfo{pages}{094502} (\bibinfo{year}{1999}), \eprint{hep-th/9704106}.

\bibitem[{\citenamefont{Wiese}(2014)}]{Wiese:2014rla}
\bibinfo{author}{\bibfnamefont{U.-J.} \bibnamefont{Wiese}},
  \bibinfo{journal}{Nucl. Phys.} \textbf{\bibinfo{volume}{A931}},
  \bibinfo{pages}{246} (\bibinfo{year}{2014}), \eprint{1409.7414}.

\end{thebibliography}

\section*{Acknowledgements}
We thank W.~Detmold, D.~B.~Kaplan, B.~Kulchytskyy, C.~Muschik, M.~J.~Savage and J.~M.~Zanotti for helpful discussions.
This work is supported in part by the U.S.~Department of Energy, Office of Science, Office of Nuclear Physics under grant Contract Number DE-SC0011090. PES is supported by the National Science Foundation under CAREER Award 1841699, and PES and AA are supported in part by Perimeter Institute for Theoretical Physics. Research at the Perimeter Institute is supported by the Government of Canada through the Department of Innovation, Science and Economic Development and by the Province of Ontario through the Ministry of Research and Innovation.
RDY is supported in part by the Australian Research Council, grant DP19010029.

\clearpage
\begin{widetext}
\section{Supplementary Material}

This supplementary material makes explicit the variational basis of interpolating operators used for the demonstration of the quantum-improved interpolating operator optimization described in the main text. A basis of 6 operators is chosen, constructed to have the appropriate parity:
\begin{align}
    (\hat{O}_1)_{\mathrm{src}}(x_i) 
        &= \hat{\ell}_i + \hat{\ell}_{i+1} \label{eq:link-op-1-2}  \\
    (\hat{O}_2)_{\mathrm{src}}(x_i) &=
       \hat{\ell}_{i} \hat{\ell}_{i-2}
        - \hat{\ell}_{i+1} \hat{\ell}_{i+3},  \label{eq:link-op-2} \\
    (\hat{O}_3)_{\mathrm{src}}(x_i) &=
        \hat{\ell}_{i} \hat{\ell}_{i-4}
        - \hat{\ell}_{i+1} \hat{\ell}_{i+5},  \label{eq:link-op-3} \\
    (\hat{O}_4)_{\mathrm{src}}(x_i) &=
        \hat{\ell}_{i} \hat{\ell}_{i-1} \hat{\ell}_{i-2}
        + \hat{\ell}_{i+1} \hat{\ell}_{i+2} \hat{\ell}_{i+3},   \label{eq:link-op-4}  \\
    (\hat{O}_5)_{\mathrm{src}}(x_i) &=
        \hat{\ell}_{i} \hat{\ell}_{i-2} \hat{\ell}_{i-4} 
        + \hat{\ell}_{i+1} \hat{\ell}_{i+3} \hat{\ell}_{i+5},     \label{eq:link-op-5} \\
    (\hat{O}_6)_{\mathrm{src}}(x_i) &=
        \hat{\ell}_{i} \hat{\ell}_{i-2} \hat{\ell}_{i-5} 
        + \hat{\ell}_{i+1} \hat{\ell}_{i+3} \hat{\ell}_{i+6},     \label{eq:link-op-6}
\end{align}
Note that the operator in Eq.~\ref{eq:link-op-1-2} may act non-trivially on configurations with one or more electron-positron pairs; operators in Eq.~\eqref{eq:link-op-2}--\eqref{eq:link-op-3}, two or more; operators in Eq.~\eqref{eq:link-op-4}--\eqref{eq:link-op-6}, three or more. 
For the more harsh truncations in $\tilde\Lambda^2$, some of these operators annihilate the entire space and hence the corresponding optimisation (Eq.~\eqref{eq:OVQE}) is performed with respect to a smaller operator basis.

\end{widetext}

\end{document}